\documentclass[prx,aps,showpacs,superscriptaddress,nofootinbib]{revtex4-1}
\usepackage[top=3cm, bottom=2cm, left=2cm, right=2cm]{geometry}
\usepackage{graphicx}
\usepackage{amsbsy}
\usepackage{amsmath}
\usepackage{xcolor}
\usepackage{amssymb}

\makeatletter
\@booleanfalse\titlepage@sw
\makeatother

\begin{document}

\title{Asymmetric comb waveguide for strong interactions between atoms and light}

\author{N. Fayard}
\address{Universit{\'e} Paris-Saclay, Institut d'Optique Graduate School, CNRS, Laboratoire Charles Fabry, 91127
Palaiseau, France}
\author{A. Bouscal}
\address{Laboratoire Kastler Brossel, Sorbonne Université, CNRS,
ENS-Universit{\'e} PSL, Collège de France, 4 place Jussieu, 75005 Paris, France}
\author{J. Berroir}
\address{Laboratoire Kastler Brossel, Sorbonne Université, CNRS,
ENS-Universit{\'e} PSL, Collège de France, 4 place Jussieu, 75005 Paris, France}
\author{A. Urvoy}
\address{Laboratoire Kastler Brossel, Sorbonne Université, CNRS,
ENS-Universit{\'e} PSL, Collège de France, 4 place Jussieu, 75005 Paris, France}
\author{T. Ray}
\address{Laboratoire Kastler Brossel, Sorbonne Université, CNRS,
ENS-Universit{\'e} PSL, Collège de France, 4 place Jussieu, 75005 Paris, France}
\author{S. Mahapatra}
\address{Centre de Nanosciences et de Nanotechnologies, CNRS, Université Paris-Saclay, 91120 Palaiseau, France}
\author{M. Kemiche}\altaffiliation[Current affiliation: ]{Univ. Grenoble Alpes, Univ. Savoie Mont Blanc, CNRS, Grenoble INP, IMEP-LAHC, 38000 Grenoble, France.}
\address{Centre de Nanosciences et de Nanotechnologies, CNRS, Université Paris-Saclay, 91120 Palaiseau, France}
\author{J. A. Levenson}
\address{Centre de Nanosciences et de Nanotechnologies, CNRS, Université Paris-Saclay, 91120 Palaiseau, France}
\author{J.-J. Greffet}
\address{Universit{\'e} Paris-Saclay, Institut d'Optique Graduate School, CNRS, Laboratoire Charles Fabry, 91127
Palaiseau, France}
\author{K. Bencheikh}
\address{Centre de Nanosciences et de Nanotechnologies, CNRS, Université Paris-Saclay, 91120 Palaiseau, France}
\author{J. Laurat}
\address{Laboratoire Kastler Brossel, Sorbonne Université, CNRS,
ENS-Universit{\'e} PSL, Collège de France, 4 place Jussieu, 75005 Paris, France}
\author{C. Sauvan}\email[Corresponding author: ]{christophe.sauvan@institutoptique.fr}
\address{Universit{\'e} Paris-Saclay, Institut d'Optique Graduate School, CNRS, Laboratoire Charles Fabry, 91127
Palaiseau, France}

\date{\today}
\begin{abstract}
Coupling quantum emitters and nanostructures, in particular cold atoms and waveguides, has recently raised a large interest due to unprecedented possibilities of engineering light-matter interactions. However, the implementation of these promising concepts has been hampered by various theoretical and experimental issues. In this work, we propose a new type of periodic dielectric waveguide that provides strong interactions between atoms and guided photons with an unusual dispersion. We design an asymmetric comb waveguide that supports a slow mode with a quartic (instead of quadratic) dispersion and an electric field that extends far into the air cladding for an optimal interaction with atoms. We compute the optical trapping potential formed with two guided modes at frequencies detuned from the atomic transition. We show that cold Rubidium atoms can be trapped as close as 100~nm from the structure in a 1.3-mK-deep potential well. For atoms trapped at this position, the emission into guided photons is largely favored, with a beta factor as high as 0.88 and a radiative decay rate into the slow mode 10 times larger than the free-space decay rate.       
\end{abstract}

\maketitle

\section{Introduction}

The development of quantum technologies requires a number of prerequisites among which strong atom-photon interactions hold a high rank~\cite{kimble2008quantum,chang2014quantum}. Enhancing the interaction between a single photon and a single atom has been a driving force for a large community over the past decades. Two main routes have first been followed rather independently: atoms in macroscopic high-finesse cavities~\cite{raimond2001manipulating,reiserer2015cavity} or solid-state emitters (e.g., quantum dots) in micro and nanostructures~\cite{GerardPRL98,lodahl2015interfacing}. In recent years, various works have combined both approaches by interfacing cold atoms with nanophotonic devices such as photonic crystal nanocavities~\cite{tiecke2014nanophotonic}, nanofibers~\cite{nayak2007optical,vetsch,goban2012demonstration}, and photonic crystal waveguides~\cite{goban2014atom,goban2015superradiance,zang2016interaction}. These hybrid strategies benefit from both the long coherence time of atoms and the enhanced electromagnetic field associated with subwavelength light confinement in nanostructures. 

Besides the well-established field of cavity quantum electrodynamics (QED), the use of single-pass schemes with nanofibers or nanostructured waveguides has triggered the emergence of a new field of research known as waveguide QED~\cite{roy2017colloquium,chang2018colloquium,turschmann2019coherent}. Photons travelling in the waveguide carry the information through long distances while atoms can store it for long times. These systems are a promising building block for quantum networks as shown recently by the experimental demonstrations of a coherent photon storage~\cite{gouraud2015demonstration,sayrin2015storage}, the heralded creation of a single collective excitation of atomic arrays~\cite{corzo2019waveguide}, or a correlated photon transport ~\cite{mahmoodian2018strongly,mahmoodian2020dynamics,prasad2020correlating}. In addition, strong atom-photon interactions give rise to new phenomena in many-body physics~\cite{douglas2015quantum} such as the emergence of solitons dynamics~\cite{calajo2021few,bakkensen2021photonic} or many-body localization~\cite{fayard2021many}.

Most of the existing experiments in waveguide QED at optical frequencies are performed with silica nanofibers~\cite{nayak2007optical,vetsch,goban2012demonstration,gouraud2015demonstration,sayrin2015storage,sorensen2016coherent,Corzo2016,solano2017super,corzo2019waveguide,prasad2020correlating}. They offer a 4$\pi$ solid-angle access to the interaction region that markedly eases the manipulation of atomic clouds near the structure. In addition, nanofibers provide single-mode operation over a broad spectral range. It is thus possible to use guided light beams at frequencies detuned from the atomic transition for trapping the atoms at subwavelength distances from the waveguide~\cite{goban2012demonstration,balykin2004atom,le2004atom}. However, the atom-photon interaction remains relatively weak. It can be quantified by the $\beta$ factor, which is defined as the ratio of the radiative decay rate of a single atom in the waveguide mode $\Gamma_{1\textrm{D}}$ to the total decay rate $\Gamma_{\textrm{tot}}$, $\beta = \Gamma_{1\textrm{D}}/\Gamma_{\textrm{tot}} = \Gamma_{1\textrm{D}} / (\Gamma_{1\textrm{D}} + \Gamma')$. Photons that are not funnelled into the waveguide mode are lost in the radiation continuum with a decay rate $\Gamma'$. Experimentally observed values of $\beta$ in nanofibers are typically in the range of $\beta \sim 0.05$~\cite{nieddu2016optical,forn2019ultrastrong,sheremet2021waveguide}.  

A promising route to increase $\Gamma_{1\textrm{D}}$ and $\beta$ is to use periodic dielectric waveguides, i.e., waveguides with a periodic modulation of the refractive index along the propagation direction. Indeed, the radiative decay rate of a single atom in a guided mode is given by $\Gamma_{1\textrm{D}} / \Gamma_0 = n_g \sigma / (2A_\textrm{eff})$, where $\Gamma_0$ is the atomic decay rate in free space, $\sigma$ is the absorption cross-section, $n_g$ and $A_\textrm{eff}$ are the group index of the mode and the effective area at the atom position~\cite{goban2014atom,lecamp2007prl}. In periodic waveguides, the coupling between contrapropagating modes results in the opening of bandgaps in the dispersion relation and in the apparition of band edges where the group velocity $v_g = c/n_g$ goes to zero~\cite{joannopoulos2008molding}. Close to these peculiar points, periodic waveguides support slow guided modes with large $n_g$'s, which, in turn, lead to increased values of $\Gamma_{1\textrm{D}}$ and $\beta$~\cite{lecamp2007prl}. Hence, an important research effort has been devoted to the design and the realization of periodic waveguides aimed at increasing the interaction with cold atoms~\cite{goban2014atom,goban2015superradiance,zang2016interaction,hung2013trapped,yu2014nanowire}. However, this is a challenging task that requires tackling various issues~\cite{zang2016interaction,kimble2020optica}, including the sensitivity of slow light to fabrication imperfections, a stable trapping of atoms at the desired positions, and the accessibility around the structure to ease the transport of the atoms to the trapping sites. 

In addition to enhanced atom-photon interactions, periodic waveguides also offer the possibility to create dispersion relations that greatly differ from the usual and almost linear dispersion of a nanofiber. The curvature of the dispersion relation between the frequency $\omega$ and the wavevector $k$ has a profound impact on various physical phenomena, such as the distortion of short pulses during propagation, the collective properties of atomic arrays coupled to the waveguide, the localization of photons emitted in the bandgap~\cite{chang2018colloquium}, as well as the tolerance of the slow mode to inevitable fabrication imperfections~\cite{zang2016interaction,FaggianiSR16}. However, the engineering of the dispersion has been quite limited until now. Indeed, most of the periodic waveguides studied so far (be it for interacting with atoms or for another purpose) are made of a periodic pattern that is symmetric in transverse directions~\cite{goban2014atom,goban2015superradiance,hung2013trapped,joannopoulos2008molding,LoncarAPL00,NotomiPRL01,DeLaRueOE09,ChebenOE10,ChebenOE12,ChebenOL14,VivienSR19} and their dispersion relation varies most often quadratically at band edges.

By breaking the transverse symmetry, one can obtain new degrees of freedom for engineering the dispersion beyond the standard parabolic shape. Recently, Nguyen \emph{et al.} demonstrated with a two-dimensional (2D) structure that symmetry breaking can be used to create exotic photonic dispersion relations such as Dirac cones, multivalley curves, or flat bands~\cite{nguyen2018symmetry}.

\begin{figure}   
\begin{center}     
\includegraphics[width=0.4\linewidth]{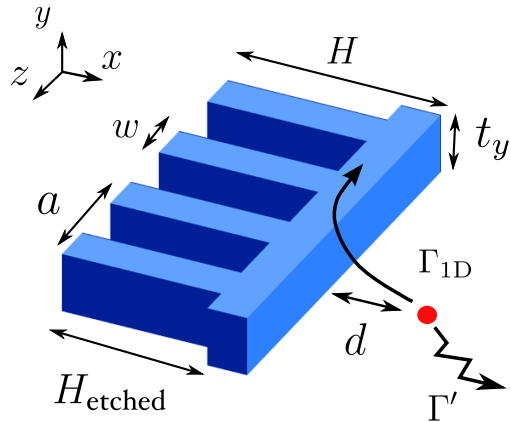} 
             \end{center}
   \caption{Scheme of the proposed asymmetric comb waveguide. The structure is etched in a GaInP membrane (refractive index $n=3.31$) suspended in air. The waveguide has a total width $H$ in the transverse $x$ direction and a thickness $t_y$ in the transverse $y$ direction. The comb pattern along the propagation direction $z$ is made of teeth with a width $w$ and a height $H_{\textrm{etched}}$, periodically spaced with a period $a$. The width of the guiding region is thus $H-H_{\textrm{etched}}$. An excited atom (red dot) decays radiatively either into the guided mode, with a rate $\Gamma_{1\textrm{D}}$, or into the radiation continuum with a rate $\Gamma'$.}
 \label{fig:structure}   
\end{figure}

In this article, we exploit symmetry breaking to design a novel three-dimensional (3D) geometry, the asymmetric comb waveguide, that (i) supports a slow mode with a \emph{quartic} dispersion, (ii) offers the possibility to trap atoms optically at subwavelength distances, and (iii) provides very large $\beta$ factors for trapped atoms. We show that the quartic dispersion makes the slow mode more tolerant to fabrication imperfections than previous proposals~\cite{zang2016interaction}. The structure is represented in Fig.~\ref{fig:structure}. It consists of a suspended bridge waveguide (with a 4$\pi$ solid-angle access to the interaction region like a nanofiber) of width $H$ and thickness $t_y$ that has been periodically corrugated with an asymmetric rectangular pattern. This promising structure that provides both a strong atom-photon interaction and a dispersion that differs from the usual parabolic shape opens new opportunities in waveguide QED.

The article is organized as follows. In Section~\ref{sec:2Dcomb}, we illustrate with 2D examples the role of symmetry breaking in the design of a flat band with a quartic dispersion. Then, we design in Section~\ref{sec:3Dcomb} a 3D comb waveguide with a quartic dispersion after having detailed a few important criteria that should be fulfilled by a periodic waveguide aimed at interacting with cold atoms. We present in Section~\ref{sec:trapping} the calculation of the potential of a two-color optical trap made with guided modes. We show that cold Rubidium (Rb) atoms can be trapped as close as 100~nm from the waveguide in a 1.3-mK-deep potential well by using relatively low powers ($P \sim 1$ mW) compatible with photonic structures. Finally, we calculate in Section~\ref{sec:interaction} the emission rates $\Gamma_\textrm{1D}$, $\Gamma_\textrm{tot}$, and $\Gamma'$ of a trapped atom. We provide evidence for very large coupling between the atom and a slow mode with $n_g = 50$: $\Gamma_\textrm{1D} = 10 \Gamma_0$, $\Gamma' = 1.3 \Gamma_0$, and $\beta = 0.88$. Section~\ref{sec:conclu} concludes the work and discusses some perspectives.

\section{Slow mode with a quartic dispersion}
\label{sec:2Dcomb}

\begin{figure*}[t]   
\begin{center}     
\includegraphics[width=0.8\textwidth]{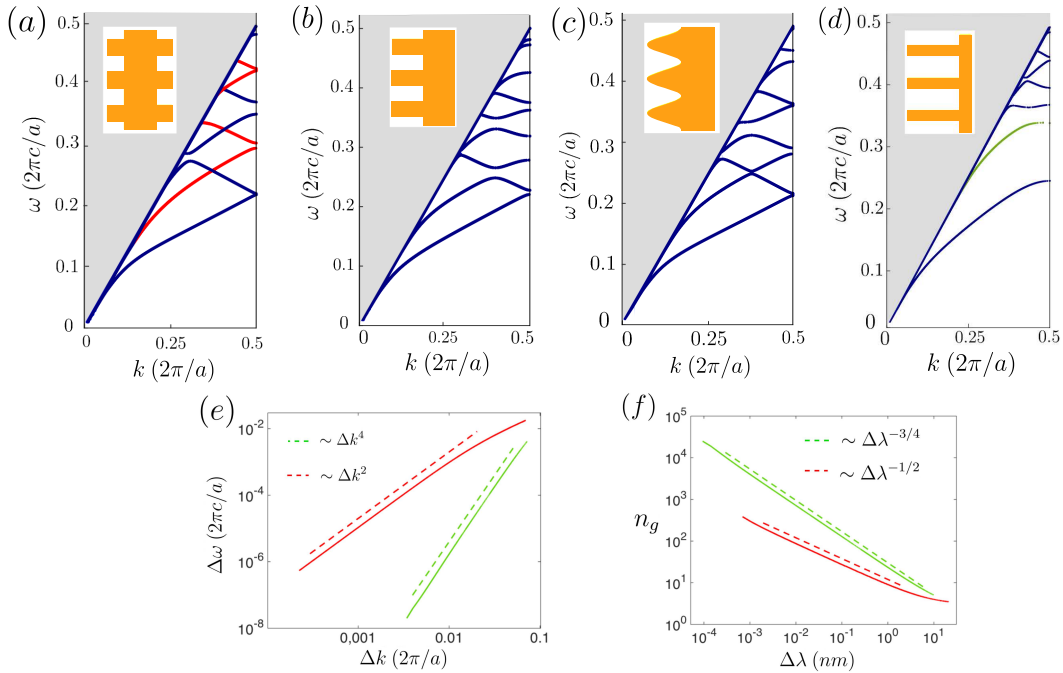} 
             \end{center}
   \caption{Role of a transverse symmetry breaking. (a)-(d) Band diagram $\omega = f(k)$ of four different 2D comb waveguides in TM polarization. The total height of all structures is $H=2a$, with $a$ the period. To bring these 2D calculations closer to 3D simulations, we consider an effective refractive index $n=2.85$. (a) Symmetric comb waveguide. The width of the teeth on both sides is $w=0.5a$ and their depth is $H_{\textrm{etched}}=0.25H$. (b) Asymmetric comb waveguide. The teeth are two times deeper, $H_\textrm{etched}=0.5H$, with the same width $w=0.5a$ to conserve the fraction of matter. (c) Asymmetric comb waveguide with a sinusoidal corrugation but the same fraction of matter. The depth of the cosine modulation is chosen so that the quantity of matter is the same as in (a)-(b). (d) Asymmetric comb waveguide designed to support a flat band with a quartic dispersion (green curve). The width of the teeth is $w=0.372a$ and their depth is $H_\textrm{etched}=0.8H$. (e) Variation of $\Delta \omega=\vert\omega-\omega_e\vert$ as a function of $\Delta k=\vert k - \pi/a\vert$ in logarithmic scales for the green band in (d) and the red band of lowest frequency in (a). We observe that $\Delta \omega\sim \Delta k^4$ for the flat band of the asymmetric comb and $\Delta \omega\sim \Delta k^2$ for the band of the symmetric comb. Those different behaviors lead to different scaling laws for the variation of the group index as a function of $\Delta \lambda=\vert \lambda-\lambda_e\vert $ as shown in (f).}
 \label{fig:symmetry}   
\end{figure*}

In this Section, we show that symmetry breaking can be used to create a comb waveguide that supports a slow mode with a quartic dispersion around a zero-group-velocity point instead of the usual parabolic shape. In order to illustrate the impact of symmetry breaking, we compare the band diagrams of four different periodic waveguides with a period $a$. The waveguides are schematically represented in the insets of Figs.~\ref{fig:symmetry}(a)-(d). All four are comb waveguides (see Fig.~\ref{fig:structure}) with the same total width $H=2a$ but different corrugations. The comparison is done with 2D structures ($t_y = \infty$) in TM polarization (magnetic field polarized in the $y$ direction). We consider a refractive index $n=2.85$ that corresponds to the effective index of the fundamental guided mode in a GaInP membrane of thickness 150 nm at $\lambda = 780$ nm. 

The dispersion curves are calculated with a Bloch-mode solver developed for studying light scattering in periodic waveguides~\cite{LecampOE07}. The solver calculates the wavevector $k$ in the propagation direction as a function of the frequency $\omega$. It is implemented with the aperiodic Fourier modal method (a-FMM)~\cite{Silberstein01}, which relies on an analytical integration of Maxwell's equations along the direction of periodicity ($z$ axis) and on a supercell approach with perfectly-matched layers (PMLs) along the transverse directions ($x$ and $y$ axis). PMLs ensure a correct treatment of far-field radiation conditions; they are implemented as complex nonlinear coordinate transforms~\cite{Hugonin05}. This numerical method is also used in the rest of the paper for the calculation of the field distributions and the decay rates. 

Figure~\ref{fig:symmetry}(a) displays the band diagram of a comb waveguide whose corrugation is symmetric in the transverse direction. It can be easily understood with the following picture: the uncorrugated planar waveguide of width $H$ supports symmetric and antisymmetric modes and the periodic corrugation couples forward and backward propagating modes, resulting in the opening of photonic bandgaps~\cite{joannopoulos2008molding}. However, the symmetry of the corrugation forbids coupling between modes of different symmetry. Therefore, the Bloch modes resulting from the coupling can be sorted in two distinct families: they are either symmetric (blue curves) or antisymmetric (red curves). As can be seen, blue and red curves cross each other without coupling. Most of the bandgaps and the associated zero-group-velocity points are located at the edge of the first Brillouin zone, at $k = \pi/a$. Using a standard coupled-mode approach with two waves having linear dispersions, one can show that the dispersion relation of the Bloch modes varies quadratically in the vicinity of these points: $\omega - \omega_e \propto \pm (k - \pi/a)^2$, where $\omega_e$ is the frequency of the band edge where $v_g=0$. 

We now consider in Figs.~\ref{fig:symmetry}(b)-(c) comb waveguides without mirror symmetry in the transverse direction and with two different corrugations profiles. For the sake of comparison, the fraction of matter (the orange area) is the same as in Fig.~\ref{fig:symmetry}(a). The situation is now fundamentally different: since the corrugations have no particular symmetry, all possible couplings are indeed allowed. As a result, gaps are now open inside the Brillouin zone ($k \neq \pi/a$) around the points of intersection between the dispersion curves of the symmetric and antisymmetric modes of the uncorrugated waveguide. Looking precisely to the band diagrams, we observe that the spectral widths of the bandgaps are smaller for a sinusoidal profile than for a rectangular profile. Thus, the latter provides larger coupling strengths. This is the reason why we study the rectangular asymmetric comb waveguide in the following.


For a rectangular profile, the coupling strengths depend on the size of the teeth, fixed by the width $w$ and the depth $H_{\textrm{etched}}$ defined in Fig.~\ref{fig:structure}. Stronger corrugations result in larger coupling strengths and wider bandgaps. If the coupling is strong enough, two consecutive bandgaps may overlap. In this case, the coupling mechanism is more complex since it involves four waves instead of two~\cite{nguyen2018symmetry}. This offers extra degrees of freedom for engineering the bands beyond the standard parabolic shape. 

Following this principle, we have designed an asymmetric comb waveguide that supports a flat band with a \emph{quartic} dispersion curve of the form $\omega - \omega_e \propto -(k - \pi/a)^4$. This flat band is highlighted in green in Fig.~\ref{fig:symmetry}(d). In order to confirm the quartic variation, we plot in Fig.~\ref{fig:symmetry}(e) $\Delta \omega = \vert\omega - \omega_e\vert$ as a function of $\Delta k = \vert k - \pi/a\vert$ with logarithmic scales. The solid green curve that is extracted from the green band in Fig.~\ref{fig:symmetry}(d) follows a clear quartic variation, as indicated by the dashed green line. In contrast, the symmetric comb waveguide of Fig.~\ref{fig:symmetry}(a) supports modes with a quadratic dispersion, as shown by the solid and dashed red curves.

A flat band with a quartic dispersion broadens the useful bandwidth of the slow mode, i.e., the bandwidth over which $n_g$ is larger than a target value. Indeed, a quartic dispersion produces a group index that scales with the wavelength as $\Delta \lambda^{-3/4}$ while the group index of a quadratic dispersion scales as $\Delta \lambda^{-1/2}$, as can be seen in Fig.~\ref{fig:symmetry}(f). Therefore, if one wants to work at a given group index, the quartic dispersion (green curve) allows an operation at a larger $\Delta\lambda$, i.e., at a frequency further from the band edge. We will see in the next Section that this increase of the useful bandwidth is important to improve the tolerance of the slow mode to fabrication imperfections.  

Let us end this Section by emphasizing that a quartic functional form for the dispersion relation of a slow mode is uncommon, in particular in the context of waveguide QED where it has not been studied so far. It is thus important to design a waveguide with a quartic dispersion and a 3D geometry compatible with the interaction with cold atoms trapped in the air cladding. We present this design in the following Section.

\section{Design of a three-dimensional asymmetric comb waveguide}
\label{sec:3Dcomb}

We first detail a few important requirements that apply to a slow-light waveguide designed to interact with cold atoms. Then, we present the design of a 3D asymmetric comb waveguide that supports a slow mode with a quartic dispersion at the transition frequency of Rb atoms and an evanescent field that extends far into the air cladding for an optimal interaction. In addition, we pay particular attention during the design process to the presence of blue- and red-detuned modes that can be used to create an optical two-color trap. 


\subsection{Design requirements}

The design of a periodic waveguide aimed at increasing the coupling with cold atoms should maximize the emission rate into the waveguide mode $\Gamma_{1\textrm{D}}$. Since the group index diverges at a band edge, the first idea is to align the transition frequency of the atom with any band edge of the photonic dispersion diagram. Unfortunately, this simple design cannot work for several reasons. 

First, slow light is very sensitive to fabrication imperfections, resulting in backscattering, radiation losses, and potentially light localization~\cite{HughesPRL05,topolancik2007experimental,mazoyer2009disorder,FaggianiSR16}. Therefore, in a practical situation, fabrication imperfections set an upper bound to the group index $n_g$ that can be reached~\cite{patterson2009disorder,garcia2010density,MazoyerOE10}. To improve the fabrication tolerance, X. Zhang \emph{et al.} proposed to use parabolic dispersion curves with large effective photon masses~\cite{zang2016interaction}. Therefore, the design should not only increase the group index, but also reduce the curvature of the dispersion relation. 

A second issue is the value of the effective mode area at the position of the atom, or equivalently the effective cross section of the photon seen by the atom. One should be careful not to lose with an increased mode area what is gained with an increased group index. Hence, the design has to find a trade-off between two opposite trends. On the one hand, we need a mode whose field extends far into the air cladding, implying that it weakly interacts with the periodic pattern. On the other hand, we need to control the group velocity and the curvature of the dispersion relation, meaning that we need a mode that strongly interacts with the periodic pattern. 

Finally, a third important challenge is to generate a stable optical trap for the atoms at subwavelength distances of the waveguide and to be able to bring the atoms inside the trap. A fully integrated trapping scheme can be achieved by using ``fast'' guided modes (i.e., with standard values of $n_g$) at frequencies detuned from the atomic transition. For instance, red- and blue-detuned modes can be used to create a two-color trap~\cite{le2004atom}. The design of the periodic waveguide should thus ensure the presence of additional modes with adequate field profiles, which spatially overlap with each other and with the slow mode.

As a whole, the design of a periodic waveguide with increased atom-photon interactions is a complex task that should meet the following criteria: 

\begin{itemize}
    \item a slow and single-mode operation at the transition frequency of the atom (large $n_g$),
    \item a large fraction of the electric field in air outside the structure (small $A_\textrm{eff}$), 
    \item a large effective photon mass for an improved robustness to fabrication imperfections,
    \item the existence of additional modes at frequencies detuned from the atomic transition for trapping the atoms optically with low powers (few mW),
    \item a clear access around the structure to ease the transport of the atoms to the trapping sites.
\end{itemize}

Up to now, two main geometries have been investigated. The first one, the alligator waveguide, is composed of a tiny 250-nm-wide air slot symmetrically surrounded by two corrugated bridge waveguides~\cite{goban2014atom,goban2015superradiance,yu2014nanowire}. This geometry fulfills the first two criteria (large $n_g$ and small $A_\textrm{eff}$) as well as the fourth one. However, it has been shown in~\cite{zang2016interaction} that the effective photon mass of the slow mode is small, meaning that alligator waveguides are very sensitive to fabrication imperfections in the slow-light regime. The fifth criterion is not met either since the atoms need to be loaded in a narrow interacting region, which constitutes a major experimental challenge~\cite{burgers2019clocked}. External trapping via side illumination has been considered~\cite{kimble2020optica,kimble2020pnas}, but a stable trapping of atoms \emph{inside} the air slot has not been demonstrated. As a result, the value of $\beta$ and the number $N$ of trapped atoms are limited to $\beta \sim 0.5$ and $N \sim 3$~\cite{goban2015superradiance,hood2016atom}. 

A different geometry has been proposed in~\cite{zang2016interaction} to address these limitations. It consists of a hybrid-clad waveguide that combines two guidance mechanisms: total internal reflection on one lateral side with a sharp sidewall and photonic bandgap on the opposite lateral side with a two-dimensional photonic crystal. The hybrid-clad waveguide provides a flatter dispersion curve than the alligator waveguide (50 times enhanced effective photon mass) and a 2$\pi$ solid-angle access to the interaction region. However, the possibility to use detuned guided beams for trapping the atoms (fourth criterion) has not been investigated yet. One the one hand, the bandgap of the photonic-crystal cladding sets important constraints on the spectral range available to place additional guided modes that could be used for trapping. On  the other hand, the photonic-crystal geometry provides various degrees of freedom for engineering the band diagram. 

In the following, we design an asymmetric comb waveguide that fulfills all five criteria. In addition, the dispersion relation of the slow mode is quartic instead of quadratic. We show that, thanks to the quartic dispersion, the slow mode of the comb waveguide is more tolerant to fabrication imperfections than the mode of the hybrid-clad waveguide proposed in~\cite{zang2016interaction} that mostly follows a standard parabolic dispersion.


\subsection{Asymetric comb waveguide}

\begin{figure*}   
\begin{center}     
\includegraphics[width=0.9\textwidth]{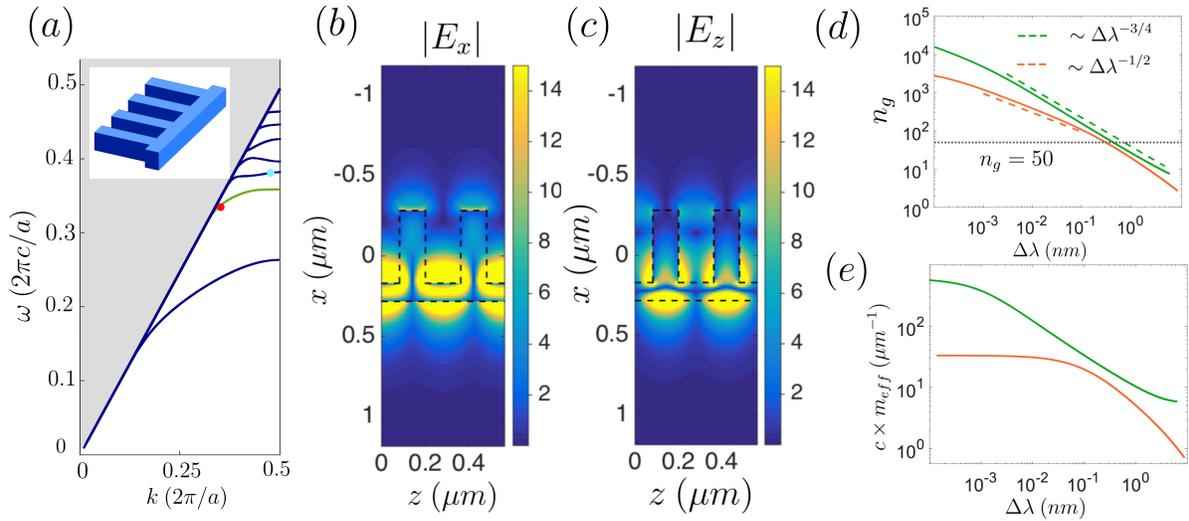} 
             \end{center}
   \caption{Slow mode with a quartic dispersion. (a) Band diagram of the 3D comb waveguide designed to support a slow mode with a quartic dispersion (green curve) that stronly interacts with Rubidium atoms. The geometrical parameters are $a=283$~nm, $H=2a$, $H_{\textrm{etched}}=0.8H$, $w=0.422a$, $t_y=150$~nm and $n=3.31$. The red and light blue dots mark the ``fast'' modes red- ($\lambda_{r}=837$ nm) and blue-detuned ($\lambda_{b}=736$ nm) with respect to the atomic transition that are used for trapping, as detailed in Section~\ref{sec:trapping}. (b) Amplitude of the $x$-component of the electric field of the slow mode. (c) Amplitude of the $z$-component of the electric field of the slow mode. The field distributions are calculated at $\lambda_0=780$ nm in the $(x,z)$ plane ($y=0$ in the middle of the membrane). (d) Group index as a function of the distance to the band edge, $\Delta\lambda = |\lambda - \lambda_e|$ (logarithmic scales). (e) Effective photon mass as a function of $\Delta\lambda = |\lambda - \lambda_e|$ (logarithmic scales). The green curves correspond to the slow mode of the 3D comb waveguide [green curve in (a)] and the orange curves correspond to the hybrid-clad waveguide in~\cite{zang2016interaction}. In (d), the dashed curves indicate the $\Delta \lambda^{-3/4}$ (green) and $\Delta \lambda^{-1/2}$ (orange) scaling laws.}
 \label{fig:Slowmode}   
\end{figure*}

Let us consider a GaInP membrane (refractive index $n = 3.31$) with a thickness $t_y = 150$~nm, as shown in Fig.~\ref{fig:structure}. Starting from the 2D structure in Fig.~\ref{fig:symmetry}(d), we have designed a 3D comb waveguide that supports a slow mode with a quartic dispersion $\omega - \omega_e \propto -(k - \pi/a)^4$ and a group index $n_g = 50$ at $\lambda_0 = 780$~nm, the wavelength of the $5S_{1/2} \leftrightarrow 5P_{3/2}$ transition of Rb atoms. The geometrical parameters are $a = 283$~nm, $H=2a$, $H_\textrm{etched}=0.8H$, and $w=0.422a$. The band diagram of this comb waveguide is shown in Fig.~\ref{fig:Slowmode}(a); the slow mode with a quartic dispersion is highlighted in green.

Figures~\ref{fig:Slowmode}(b) and~\ref{fig:Slowmode}(c) display the distribution in the $(x,z)$ plane of the dominant components of the electric field ($E_x$ and $E_z$) at $\lambda_0 = 780$~nm. Note that the evanescent tail of the electric field extends far in the air cladding. On the side opposite the teeth, the longitudinal electric-field component $E_z$ is maximum in front of the teeth, at $z \equiv a/2$~(mod $a$). The positions of the maxima of the transverse electric-field component $E_x$ are shifted by $a/2$, at $z \equiv 0$~(mod $a$). 

In addition to the slow mode, the band diagram should also provide two additional modes, red- and blue-detuned with respect to the atomic transition, that can be used to realize an optical two-color trap. These modes should lie below the light line and below the frequency of the electronic bandgap of GaInP ($\lambda=680$~nm). The design of the structure has been performed by taking into account these additional constraints. The red- and blue-detuned modes are highlighted by the red and light blue dots in Fig.~\ref{fig:Slowmode}(a). The calculation of the trapping potential is presented in Section~\ref{sec:trapping}.

To ease the future implementation of the structure, it is important to keep in mind that nanostructured devices are never manufactured to their nominal specifications. Hence, real periodic waveguides are not exactly periodic and suffer from random fabrication imperfections. Slow light is particularly sensitive to these imperfections. It experiences random scattering as it propagates, resulting in backscattering, radiation losses, and possibly light localization~\cite{HughesPRL05,topolancik2007experimental,mazoyer2009disorder}. 

In the usual case of a parabolic dispersion, it has been shown in~\cite{zang2016interaction,FaggianiSR16} that the important physical parameter to be considered to evaluate the sensitivity of slow light to fabrication imperfections is the effective photon mass, $m_\textrm{eff} = \left ( \partial^2\omega/\partial k^2 \right ) ^{-1}$. It is a constant for a parabolic dispersion. X. Zhang \emph{et al.} concluded that one should consider a dispersion curve with a large effective photon mass to improve the fabrication tolerance~\cite{zang2016interaction}.

However, since the dispersion curve of the asymmetric comb waveguide is not parabolic, we cannot use this unique figure of merit to assess the tolerance of the slow mode to fabrication imperfections. Therefore, we have extended the reasoning in~\cite{zang2016interaction} to a general, non-parabolic, dispersion. In this case, one must consider two different physical parameters: the variation of the group velocity with the frequency, $\partial v_g / \partial\omega$, and the distance of the operation frequency to the band edge, $\Delta\omega = |\omega - \omega_e|$. The first parameter $\partial v_g / \partial\omega$ is inversely proportional to the effective photon mass, which is not a constant in the case of a non-parabolic dispersion. On the other hand, the second parameter $\Delta\omega = |\omega - \omega_e|$ is independent of $m_\textrm{eff}$ if the dispersion is not parabolic. To improve the fabrication tolerance, one has to increase both parameters: the effective photon mass \emph{at the operation frequency} and the distance of the operation frequency to the band edge.

We have calculated these figures of merit for the asymmetric comb. They are shown by the green curves in Figs.~\ref{fig:Slowmode}(d) and~\ref{fig:Slowmode}(e). For the sake of comparison, we have also calculated the same parameters for the hybrid-clad waveguide proposed in~\cite{zang2016interaction}, which is the most resilient to fabrication imperfections in the waveguide QED literature. Figure~\ref{fig:Slowmode}(d) displays the distance of the operation wavelength to the band edge, $\Delta\lambda = |\lambda - \lambda_e|$, as a function of the group index. Regardless of the value of the group index, the comb waveguide allows one to operate further from the band edge than the hybrid-clad waveguide. The dashed green and orange curves show that the asymmetric comb waveguide follows the scaling law that is expected for a quartic dispersion while the hybrid-clad waveguide mostly follows the scaling law of a quadratic dispersion. The dashed horizontal line marks the value $n_g=50$ that is used in the following. Figure~\ref{fig:Slowmode}(e) shows the effective photon mass as a function of $\Delta\lambda$. Regardless of the operation frequency, the asymmetric comb has a larger effective mass than the hybrid-clad waveguide. These results allow us to conclude that the asymmetric comb waveguide is more tolerant to fabrication imperfections than the hybrid-clad waveguide in~\cite{zang2016interaction}.

\section{Two-color optical trap for Rubidium atoms}
\label{sec:trapping}

We show hereafter that the versatility of the asymmetric comb allows us to trap Rb atoms optically as close as 100 nm from the waveguide. At these deeply subwavelength distances, the atoms can strongly interact with the slow mode aligned with their transition frequency. The objective is to trap atoms where the electric field of the slow mode is intense, see Figs.~\ref{fig:Slowmode}(b)-(c). For accessibility reasons, it is not appropriate to create a trap between the comb teeth. Therefore, we aim at trapping atoms on the side opposite the teeth, see the red dot in Fig.~\ref{fig:structure}.  

There exist various ways to trap atoms close to a dielectric structure~\cite{balykin2004atom}. In waveguide QED, a smart, fully integrated, approach consists in taking advantage of the waveguide to build an optical trap with guided light beams. In this work, we use a two-color trap, whose principle is to send two additional light beams into the waveguide~\cite{le2004atom}. The first one is red detuned with respect to the atomic transition; it produces a negative Stark shift on the energy of the ground state $5S_{1/2}$ that results in an attracting potential $U_r(x,y,z) < 0$. The second beam is blue detuned; it creates a positive Stark shift and a repulsive potential $U_b(x,y,z) > 0$. The combination of the two potentials,

\begin{equation}
    U_\textrm{trap}(x,y,z) = U_r(x,y,z) + U_b(x,y,z) ,
\end{equation}

\noindent can produce a deep potential well (i.e., a stable trap) at a given position in space that depends on the spatial profile of both guided beams, on their powers, and on the polarizability of the atom.

On top of the trapping potential $U_\textrm{trap}$ generated by the detuned beams, it is necessary to take into account the attracting Casimir Polder (CP) potential that arises when an atom interacts with electromagnetic vacuum fluctuations near a dielectric surface~\cite{casimir1948influence,buhmann2007dispersion}. We assume that the CP potential felt by an atom close to the vertical sidewall of the comb (see the red dot in Fig.~\ref{fig:structure}) is the same as the potential near a planar surface, which varies as 

\begin{equation}
    U_{\textrm{CP}} = -\frac{C_3}{d^3} ,
\end{equation}

\noindent where $d$ is the distance between the atom and the surface and $C_3$ is a constant that depends both on the material and the atom. The value of $C_3$ for GaInP and Rb atoms, $C_3 \simeq 6.7 \times 10^4$~mK.nm$^3$, is computed with the formula in~\cite{caride2005dependences}; the permittivity of GaInP is taken from the experimental data in~\cite{schubert1995optical}. The desired depth of the trapping potential for cold atoms is typically of the order of 1~mK. The CP potential is thus dominant close to the surface and it becomes negligible for distances larger than $100$~nm: $U_{\textrm{CP}}(10~\textrm{nm}) \sim -67$~mK and $U_{\textrm{CP}}(100~\textrm{nm}) \sim -6.7\times 10^{-2}$~mK. The CP potential makes it very difficult to trap atoms stably at distances well below 100~nm.  

\begin{figure*}   
\begin{center}     
\includegraphics[width=0.8\linewidth]{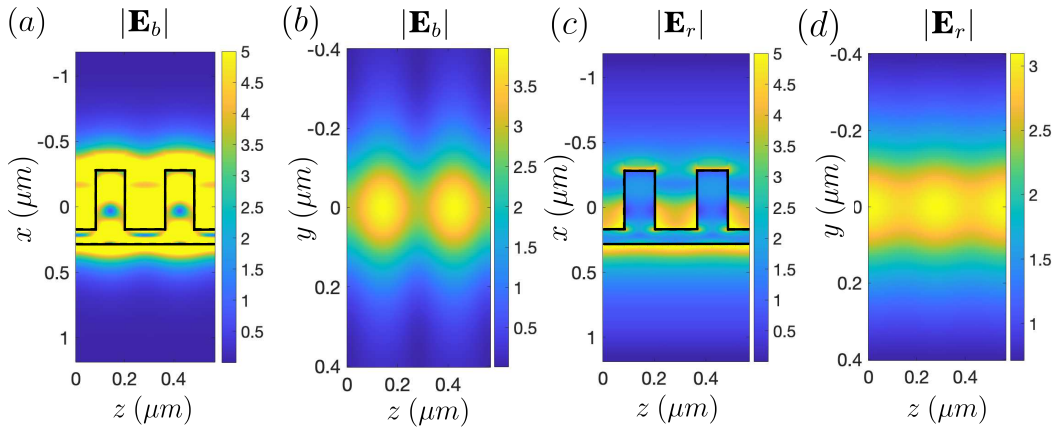} 
\end{center}
   \caption{Electric-field distributions of the blue- and the red-detuned guided modes used for the two-color trap. (a)-(b) Blue-detuned mode $\mathbf{E}_b$ at $\lambda_{b}=736$ nm, marked by a light blue dot in the band diagram of Fig.~\ref{fig:Slowmode}(a). (c)-(d) Red-detuned mode $\mathbf{E}_r$ at $\lambda_{r}=837$ nm, marked by a red dot in the band diagram of Fig.~\ref{fig:Slowmode}(a). In (a) and (c) the field is represented in the $(x,z)$ plane at $y=0$ (cross-section through the center of the GaInP membrane). The solid lines show the contour of the comb waveguide. In (b) and (d) the field is represented in a $(y,z)$ plane located in air at a distance $d=100$ nm from the structure.}
 \label{fig:fieldsfortrapping}   
\end{figure*}


In order to realize a two-color trap with guided modes, the frequencies of both detuned signals have to be chosen with the band diagram in Fig.~\ref{fig:Slowmode}(a). Let us emphasize that, since the waveguide is periodic, the guided modes are Bloch modes and their field distributions are periodically modulated along the $z$ direction. It is thus possible to create a periodic array of potential wells in the $z$ direction without using contra-propagating beams. It is convenient to choose the red-detuned frequency on the same band as the slow mode (green band), but closer to the light line where the group velocity is of the order of $c/n$. Then, one of the higher-order bands can be used for the blue-detuned frequency. 

Since the wavelength of the $5S_{1/2} \leftrightarrow 5P_{3/2}$ transition of Rb atoms is $\lambda_0 = 780$ nm, we have chosen to work with $\lambda_r = 837$~nm for the red-detuned field and $\lambda_b = 736$~nm for the blue-detuned field. These wavelengths are displayed in Fig.~\ref{fig:Slowmode}(a) with red and light blue dots. The electric fields of the corresponding guided modes are shown in Fig.~\ref{fig:fieldsfortrapping}. The amplitude of the red-detuned mode is almost homogeneous in air along the $z$ direction while the amplitude of the blue-detuned mode exhibits a more pronounced periodic modulation, with intensity maxima in front of the comb teeth at $z \equiv a/2$~(mod $a$). With these two fields, an atom near the back of the comb will ``feel'' an attracting potential that is almost independent of $z$ and a repulsive potential that is stronger in front of the teeth. By playing with the relative powers of both detuned beams, it is thus possible to create a potential well with a miniumum at $z \equiv 0$~(mod $a$). At this position, the transverse component of the electric field of the slow mode is maximum, see Fig.~\ref{fig:Slowmode}(b). Therefore, an atom trapped at this position will interact efficiently with the slow mode. 

\begin{figure}   
\begin{center}     
\includegraphics[width=0.65\linewidth]{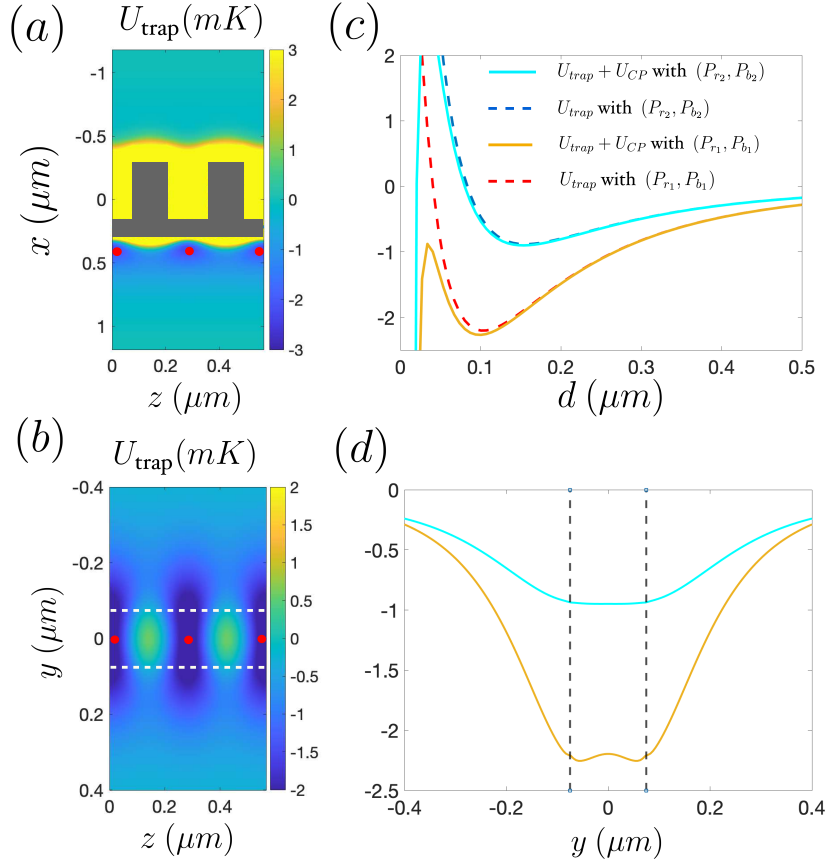} 
             \end{center}
   \caption{ Calculated potential of the two-color trap. (a)-(b) Distribution of the trapping potential $U_\textrm{trap}$ created by the superposition of the blue- and red-detuned modes. As in Fig.~\ref{fig:fieldsfortrapping}, we represent a cross-section in the $(x,z)$ plane at $y=0$ (a) and a cross-section in a $(y,z)$ plane located in air at a distance $d=100$~nm from the waveguide (b). Red dots represent atoms trapped in the potential wells. The gray area in (a) shows the comb waveguide, where the potential is not defined. The dashed lines in (b) remind the position of the GaInP membrane that is located in a different plane, shifted by 100~nm in the $x$ direction. (c) Trapping potential $U_\textrm{trap}$ (dashed curves) and total potential $U_\textrm{trap} + U_\textrm{CP}$ (solid curves) as a function of the distance $d$ to the waveguide for $y=0$ and $z=0$ (mod $a$). Two different sets of trapping powers are represented. The powers $(P_{r_1},P_{b_1}) = (1.6,1.3)$~mW create a potential well at $d=100$~nm with a depth $U \approx -2.2$~mK. The powers $(P_{r_2},P_{b_2}) = (1,1.3)$~mW create a potential well at $d=160$~nm with a depth $U \approx -0.8$~mK. (d) Total potential $U_\textrm{trap} + U_\textrm{CP}$ as a function of the vertical coordinate $y$ for $z=0$ (mod $a$). The orange (resp. light blue) curve shows the potential created at a distance $d=100$~nm (resp. $d=160$~nm) by blue- and red-detuned modes with powers $(P_{r_1},P_{b_1})$ [resp. $(P_{r_2},P_{b_2})$].}
 \label{fig:Trap}   
\end{figure}

To be quantitative, we have computed the trapping potential $U_\textrm{trap}(x,y,z)$ for Rb atoms with the recently-introduced open-source package Nanotrappy~\cite{berroir2022nanotrappy}. The inputs of Nanotrappy are the electric fields of the blue- and red-detuned guided modes, which are calculated with the mode solver for periodic waveguides used to calculate the band diagram~\cite{LecampOE07}. Nanotrappy calculates the Stark shifts induced by these electric fields and the corresponding potentials. The calculation includes the scalar, vector, and tensor shifts~\cite{berroir2022nanotrappy}.  

Figures~\ref{fig:Trap}(a)-(b) display maps of the trapping potential $U_\textrm{trap}$ in the $(x,z)$ plane at $y=0$ and in a $(y,z)$ plane located in air at a distance $d = 100$~nm from the structure. The powers of the red- and blue-detuned modes are respectively $(P_{r_1},P_{b_1}) = (1.6,1.3)$~mW. The potential is given in mK so that it can be easily compared with the temperature $T$ of cold Rb atoms in a typical experiment, $T \sim 10~\mu$K. Red dots highlight the potential minima, where atoms can be trapped, at $y=0$, $z=0$ (mod $a=283$~nm), and a distance $d = 100$~nm from the structure. The confinement is good in all three directions of space. 

Finally, we investigate in Figs.~\ref{fig:Trap}(c)-(d) the impact on the trap of the powers of the detuned modes, as well as the effect of taking into account the CP potential. Cross-sections of the trapping potential $U_\textrm{trap}$ (dashed curves) and of the total potential $U_\textrm{trap} + U_\textrm{CP}$ (solid curves) are plotted for two different sets of trapping powers $(P_{r_1},P_{b_1}) = (1.6,1.3)$~mW and $(P_{r_2},P_{b_2}) = (1,1.3)$~mW. Note that the power of the blue-detuned mode (i.e., the repulsive potential $U_b$) is kept constant. Decreasing the power of the red-detuned mode has two main effects. First, the depth of the potential well is reduced from $\approx 2.2$~mK to $\approx 0.8$~mK. Second, the trapping distance is increased from $100$~nm to $160$~nm. At such distances, adding the CP contribution does not modify significantly the potential well. It is only modified at shorter distances where the attracting CP interactions become dominant. For the closest trap ($d=100$~nm, orange solid curve), the potential barrier between the trap and the structure is lowered by the CP interactions but it remains higher than 1~mK. The orange curve in Fig.~\ref{fig:Trap}(d) shows that the potential minimum is not necessarily located at $y=0$. The height of the tiny potential barrier at $y=0$ is only 50~$\mu$K

We have demonstrated the possibility to trap cold Rb atoms at deeply subwavelength distances ($d = 100$~nm) from the comb waveguide where the $x$-component of the electric field of the slow mode is intense. We have used a two-color optical trap with powers of about 1~mW, a value well compatible with the powers used in current integrated-optics experiments. We show in the following Section that an atom trapped at this position interacts strongly with the slow mode.

\section{Strong atom-photon interaction}
\label{sec:interaction}

Having established the possibility to trap Rb atoms close to the comb waveguide, we now show that trapped atoms can strongly interact with the slow mode. For that purpose, we compute the spontaneous emission rate of an excited atom and the $\beta$ factor --~i.e., the fraction of light coupled to the slow mode.  

With the optical trap discussed in Section~\ref{sec:trapping}, atoms are located at positions where the $x$-component of the electric field of the slow mode is dominant. Therefore, we consider in the following spontaneous emission of an atom with a dipole $\mathbf{d}$ that is linearly polarized along the $x$ direction, $\mathbf{d} = d \mathbf{e}_x$ with $\mathbf{e}_x$ the unitary vector along the $x$ direction. Experimentally, this situation can be realized by imposing a quantization axis along $x$ with an external magnetic field.

We calculate the total emission rate $\Gamma_{\textrm{tot}}$ and the emission rate in the slow mode $\Gamma_{1\textrm{D}}$ (sum of the forward and backward propagation directions). The former is proportional to the imaginary part of the total Green tensor at the position $\mathbf{r}_0$ of the atom~\cite{Novotny}, 

\begin{equation}\label{eq:GammaTot}
    \frac{\Gamma_{\textrm{tot}}}{\Gamma_0} = \frac{6\pi c}{\omega_0} \mathbf{e}_x \cdot \textrm{Im}\left [ \mathbf{G}(\mathbf{r}_0,\mathbf{r}_0,\omega_0) \right ] \mathbf{e}_x ,
\end{equation}

\noindent while the latter depends only on the electric field of the guided mode~\cite{goban2014atom,lecamp2007prl}, 

\begin{equation}
    \frac{\Gamma_{1\textrm{D}}}{\Gamma_0} =  \frac{n_g \sigma a \varepsilon_0 |\mathbf{e}_x \cdot \mathbf{E}_m^\ast(\mathbf{r}_0)|^2}{2 \iiint_\textrm{Cell} \varepsilon |\mathbf{E}_m|^2 d^3\mathbf{r}} \,.
\end{equation}

\noindent In these expressions, $\omega_0$ is the transition frequency, $\Gamma_0 = \omega_0^3 |\mathbf{d}|^2 / (3\pi \hbar \varepsilon_0 c^3)$ is the emission rate in vacuum, $\sigma = 3\lambda_0^2/(2\pi)$ is the absorption cross-section, $\mathbf{E}_m$ is the electric field of the slow mode, and $n_g = c/v_g$ is its group index. One usually defines the mode effective area as $A_\textrm{eff}(\mathbf{r}_0) = \iiint_\textrm{Cell} \varepsilon |\mathbf{E}_m|^2 d^3\mathbf{r} / (a \varepsilon_0 |\mathbf{e}_x \cdot \mathbf{E}_m^\ast(\mathbf{r}_0)|^2)$. Note that the volume integral runs over one unit cell of the periodic waveguide. 

Then, we deduce the emission rate outside the slow mode (i.e., into all the other modes), $\Gamma' = \Gamma_{\textrm{tot}} - \Gamma_{1\textrm{D}}$, as well as the value of the $\beta$ factor, $\beta = \Gamma_{1\textrm{D}} / \Gamma_{\textrm{tot}}$. Let us recall that the slow mode supported by the comb waveguide designed in Section~\ref{sec:3Dcomb} has a group index $n_g = 50$ at the transition wavelength of Rb atoms, $\lambda_0 = 2\pi c/\omega_0 = 780$ nm. Its electric field is represented in Figs.~\ref{fig:Slowmode}(b)-(c).

The main difficulty is the calculation of the total Green tensor $\mathbf{G}(\mathbf{r}_0,\mathbf{r}_0,\omega_0)$ of an infinitely long periodic waveguide, which requires an accurate calculation of the emission into radiation modes with outgoing-wave conditions in a periodic medium~\cite{LecampOE07}. Two different approaches are often used to bypass this difficulty. The first one amounts to assume that the emission rate into radiation modes is approximately equal to the emission rate in vacuum, $\Gamma' \approx \Gamma_0$, see for instance~\cite{zang2016interaction}. With this approximation, Eq.~(\ref{eq:GammaTot}) simply becomes $\Gamma_{\textrm{tot}}/\Gamma_0 \approx 1 + \Gamma_{1\textrm{D}}/\Gamma_0$. In that case, the calculation only requires the knowledge of the guided mode, which can be calculated with a Bloch-mode solver. The second approach that avoids the calculation of the Green tensor of an infinitely long periodic waveguide consists in considering a finite-size structure formed by a finite number of periods, see for instance~\cite{hung2013trapped}. The main drawback is that the calculated structure (a Fabry-Perot cavity) is different from the desired one (an infinitely long waveguide). As a result, the emission rate exhibits a series of spurious resonance peaks that depends on the arbitrary choice of the structure length and termination. It is thus difficult to infer the actual emission rate of the periodic waveguide.   

\begin{figure*}   
\begin{center}     
\includegraphics[width=0.9\textwidth]{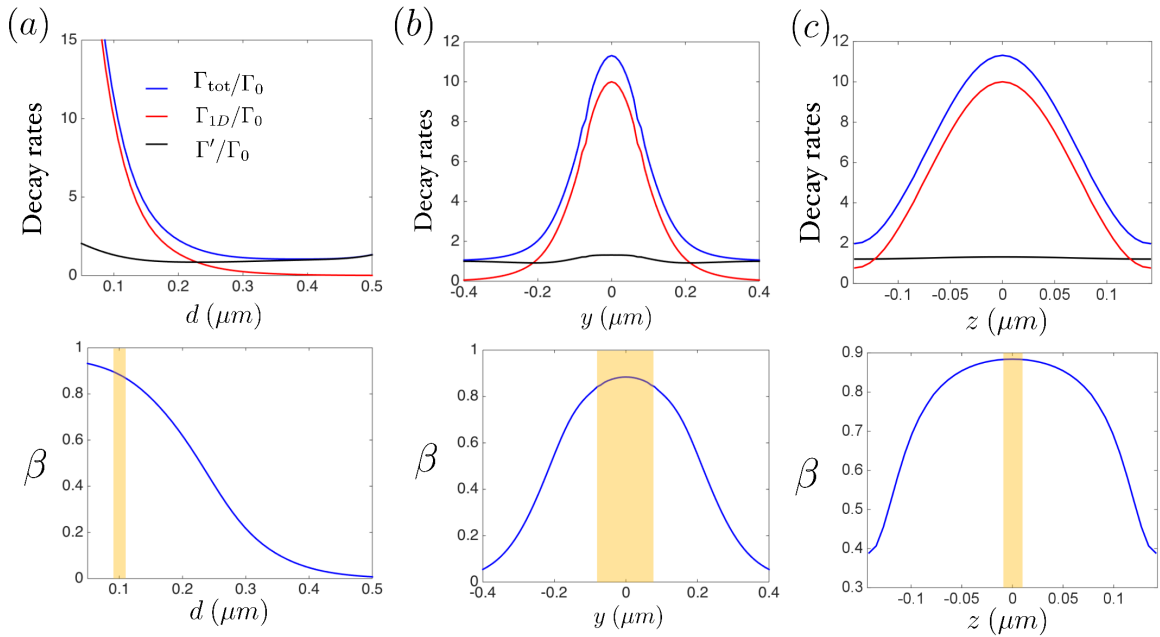} 
             \end{center}
   \caption{Decay rates of an atom near the 3D asymmetric comb waveguide. First line: Decay rates $\Gamma_\textrm{tot}$ (blue), $\Gamma_\textrm{1D}$ (red), and $\Gamma'$ (black) as a function of the distance $d$ from the waveguide for $y=0$ and $z=0$ (mod $a$) (a), as a function of $y$ for $d=100$~nm and $z=0$ (mod $a$) (b), and as a function of $z$ for $d=100$~nm and $y=0$ (c). All decay rates are normalized by the decay rate  $\Gamma_0$ of the atom in vacuum. Second line: Variation of the $\beta$ factor, $\beta = \Gamma_\textrm{1D}/\Gamma_\textrm{tot}$, at the same positions. The yellow areas represent the volume where atoms are likely to be trapped, defined as the positions where the value of the orange potential in Fig.~\ref{fig:Trap} is between $U_{min} \approx -2.2$~mK and $U_{min} + 50$~$\mu$K.}
 \label{fig:Decay}   
\end{figure*}

In contrast to these two approximate approaches, we calculate rigorously the Green tensor of the periodic waveguide by using a modal method that relies on an exact Bloch-mode expansion~\cite{LecampOE07}. Figure~\ref{fig:Decay} shows the variation of the decay rates $\Gamma_\textrm{tot}$, $\Gamma_\textrm{1D}$, $\Gamma'$ (first line), and of the $\beta$ factor (second line) as a function of the position of the atom. In Fig.~\ref{fig:Decay}(a), the atom is moved horizontally away from the comb for $y=0$ and $z=0$ (mod $a$). In Fig.~\ref{fig:Decay}(b), the atom is moved vertically for $d=100$~nm and $z=0$ (mod $a$). In Fig.~\ref{fig:Decay}(c), the atom is moved along the waveguide for $d = 100$~nm and $y=0$. The yellow areas represent the volume where atoms are likely to be trapped, defined as the positions where the value of the orange potential in Fig.~\ref{fig:Trap} is between $U_{min} \approx -2.2$~mK and $U_{min} + 50$~$\mu$K. This defines a trapping volume of typical size 30~nm along $x$, 160~nm along $y$ and 20~nm along $z$.

For large distances, $d > 400$~nm, the emission rate $\Gamma_\textrm{1D}$ into the slow mode is negligible compared to the emission rate $\Gamma'$ into the radiation continuum and the $\beta$ factor tends towards zero. As the distance $d$ decreases, the atom enters the region where the field of the slow mode is intense and Figs.~\ref{fig:Decay}(a) shows a strong enhancement of $\Gamma_\textrm{1D}$ that results in an increase of the $\beta$ factor. For a trapping distance of $d=100$~nm (yellow area), $\Gamma_\textrm{1D} = 10\Gamma_0$, $\Gamma' = 1.3\Gamma_0$, and $\beta = 0.88$. This number is significantly larger than the value of $\beta \approx 0.5$ that has been experimentally observed for the alligator waveguide~\cite{goban2015superradiance,hood2016atom}.

In Figs.~\ref{fig:Decay}(b)-(c), we fix the distance $d=100$~nm and we vary the position of the atom in the two other directions. Along the vertical direction, as long as the atom is located inside the trap in front of the structure ($-75~\textrm{nm} \leqslant y \leqslant 75~\textrm{nm}$), the decay rate into the slow mode remains dominant $\Gamma_\textrm{1D} > 8 \Gamma_0$ and $\beta > 0.8$. Along the $z$ direction, $\Gamma_\textrm{1D}$ varies periodically. Its variation over one period is directly related to the variation of $\vert E_x(z) \vert$ represented in Fig.~\ref{fig:Slowmode}(b). On the other hand, $\Gamma'$ is almost constant.

Finally, let us emphasize that the presence of the comb increases the emission rate into the radiation continuum $\Gamma'$, compared to the emission rate in vacuum $\Gamma_0$. Indeed, for $d \leqslant 100$~nm and $-75~\textrm{nm} \leqslant y \leqslant 75~\textrm{nm}$, $\Gamma' \geqslant 1.3\Gamma_0$. Therefore, the assumption $\Gamma' \approx \Gamma_0$ would lead to an error larger than $30\%$ for $d \leqslant 100$~nm. This trend of an increase of $\Gamma'$ close to a dielectric structure is not general; a previous calculation with the same numerical method has shown a decrease of $\Gamma'$ in the near field of a different periodic waveguide~\cite{LalanneOL2015}.

\section{Conclusion}
\label{sec:conclu}

We have proposed a new waveguide geometry, the asymmetric comb, that provides a strong interaction between trapped atoms and guided photons. An important originality of the structure is its quartic dispersion relation of the form $\omega - \omega_e \propto -(k - \pi/a)^4$, unique in the context of waveguide QED. We have demonstrated that this specific form of the dispersion relation reduces the impact of the inevitable fabrication imperfections. Then, we have shown how cold Rubidium atoms can be trapped at subwavelength distances ($d=100$~nm) from the structure by implementing a two-color optical trap with guided modes that are red and blue detuned with respect to the atomic transition frequency. Finally, we have completely characterized the decay of an excited atom in this complex photonic environment by calculating rigorously the decay rate $\Gamma_{1\textrm{D}}$ into the guided mode as well as the decay rate $\Gamma'$ into all other radiative channels. Atoms inside the trap decay preferentially into the slow mode (group velocity $v_g = c/50$) with a $\beta$ factor as high as 0.88. 

We have conducted preliminary studies (not shown here) that show that the fabrication of the structure is completely feasible. We have considered throughout the paper a structure made of GaInP, a semiconductor material with a high refractive index. This choice is not critical for the design. We have checked that an asymmetric comb waveguide with a quartic dispersion and a large $\beta$ factor can also be designed in a material with a lower refractive index such as silicon nitride or silicon oxide.    

Being able to control the emission with figures of merit as large as $\Gamma_{1\textrm{D}} = 10 \Gamma_0$, $\Gamma' = 1.3\Gamma_0$, and $\beta = 0.88$ puts the strong coupling regime of waveguide QED ($\Gamma_{1\textrm{D}}/\Gamma'\gg1$) within reach. In this important regime, both collective ~\cite{plankensteiner2015selective,asenjo2017exponential,kornovan2019extremely,shlesinger2021time,ferioli2021storage,reitz2022cooperative} and non-linear quantum~\cite{chang2014quantum} phenomena are enhanced, meaning that single photon switches~\cite{chang2007single} or coherent photon  storage~\cite{gouraud2015demonstration,sayrin2015storage,solano2017super} could be achieved with high efficiencies. Exotic many-body phenomena such as correlated photon transport~\cite{mahmoodian2020dynamics}, many-body localization~\cite{fayard2021many}, or fermionization of the multiple excited states~\cite{albrecht2019subradiant} could also be explored with the asymmetric comb. Finally, let us emphasize that the uncommon dispersion relation of the slow mode could dramatically alter the nature and the range of the photon-mediated interactions for atoms whose transition frequencies lie inside or outside the bandgap~\cite{gonzalez2015subwavelength,douglas2015quantum,zhang2020subradiant}. Enlarging the bandwidth of the structure also allows one to probe with reduced distortion the fast dynamics of superradiant emission~\cite{gross1982superradiance}.


\section*{Acknowledgements}
This work was supported by the French National Research Agency (ANR) under the NanoStrong Project (ANR-18-CE47-0008) and the Région Île-de-France (DIM
SIRTEQ). This project has also received funding from the European
Union’s Horizon 2020 research and innovation programme under Grant
Agreement No. 899275 (DAALI project). A.U. was supported by the European
Union (Marie Curie Fellowship SinglePass 101030421).

\bibliographystyle{naturemag}

\bibliography{biblioComb}

\end{document}